\documentclass[a4paper,prd,showpacs,noshowkeys,twocolumn]{revtex4}
\usepackage[latin1]{inputenc}
\usepackage{amsmath}
\usepackage{amssymb}
\usepackage{mathrsfs}

\providecommand{\bbA}{\mathbb{A}}

\providecommand{\tp}{\mss{\mathsf{T}}}
\providecommand{\diag}{\mathrm{diag}}
\def\Tr{\mathrm{Tr}}
\providecommand{\n}{\ensuremath{\ms{N}}}
\providecommand{\nn}{\ensuremath{\mt{N}}}
\providecommand{\lc}{\ensuremath{LC^{\uparrow}}}
\providecommand{\mink}{\ensuremath{\mathbb{R}^{1,d}}}
\providecommand{\hs}[1]{\hspace{#1}}
\providecommand{\mn}[1]{\mbox{\normalsize $#1$}}
\providecommand{\mss}[1]{\mbox{\scriptsize $#1$}}
\providecommand{\ms}[1]{\mbox{\small $#1$}}

\providecommand{\mt}[1]{\mbox{\tiny $#1$}}
\providecommand{\eq}[1]{\begin{equation} #1 \end{equation}}
\providecommand{\eqarr}[1]{\begin{eqnarray} #1 \end{eqnarray}}
\providecommand{\aver}[1]{\langle #1 \rangle}
\providecommand{\re}{\mathrm{Re}}
\providecommand{\im}{\mathrm{Im}}
\def\pr{\mbox{\tiny $\prime$}}

\font\bb=bbmss12 scaled 1000
\def\id{\mbox{\bb 1}}
\begin{document}
%%%%%%%%%%%%%%%%%%%%%%%%%%%%%%%%%%%%%%%%%%%%%%%%%
\title{
The structure of potentials with $N$ Higgs doublets
}
\author{C.~C.~Nishi}
\email{ccnishi@ift.unesp.br}
\affiliation{
Instituto de F\'\i sica Te\'orica,
UNESP -- S\~ao Paulo State University\\
Rua Pamplona, 145,
01405-900 -- S\~ao Paulo, Brasil
}
\affiliation{
Instituto de Física Gleb Wataghin, UNICAMP,\\
PO Box 6165, 13083-970, Campinas, SP, Brasil}

%\date{\today}
%%%%%%%%%%%%%%%%%%%%%%%%%%%%%%%%%%%%%%%%%%%%%%%%%
\begin{abstract}
Extensions of the Standard Model with $N$ Higgs doublets are simple extensions
presenting a rich mathematical structure. An underlying Minkowski structure
emerges from the study of both variable space and parameter space. The former
can be completely parametrized in terms of two future lightlike Minkowski
vectors with spatial parts forming an angle whose cosine is
$\ms{-(N\!-\!1)^{-1}}$. For the parameter space, the Minkowski parametrization
enables one to impose sufficient conditions for bounded below potentials,
characterize certain classes of local minima and distinguish charge breaking
vacua from neutral vacua. A particular class of neutral minima presents a
degenerate mass spectrum for the physical charged Higgs bosons.

\end{abstract}
%%%%%%%%%%%%%%%%%%%%%%%%%%%%%%%%%%%%%%%%%%%%%%%%%
%\keywords{ }
%\twocolumn
\maketitle
%%%%%%%%%%%%%%%%%%%%%%%%%%%%%%%%%%%%%%%%%%%%%%%%%
\section{Introduction}
\label{sec:intro}

The scalar sector of the Standard Model (SM) is the only directly untested part
of this successful model which accounts for all the variety of phenomena
involving subnuclear particles\,\cite{pdg}. The proper knowledge of the only
elementary scalar in the SM, the Higgs, is critically important to test one of
the major features of the SM, the Higgs mechanism, responsible to give masses to
all massive gauge bosons and fermions and to hide the $SU(2)_L\otimes U(1)_Y$
symmetry\,\cite{quigg:07}. The discovery of the Higgs is eagerly awaited to
happen in the LHC experiment\,\cite{quigg:07}.

Several theoretical reasons, however, force us to consider the possibility of
more than one elementary scalar\,\cite{haber:book,CDF:07,binoth:06}. One of the
reasons is the increasingly accepted notion that the SM is possibly a low energy
manifestation of a more fundamental, yet unknown, theory such as Grand Unified
theories, with or without supersymmetry, or extra-dimensional theories, which
contain more scalars in general\,\cite{murayama:03}. The search for physics
beyond the SM is well motivated by several theoretical incompleteness features
or problems the SM faces\,\cite{murayama:03}. For example, the minimal
supersymmetric SM (MSSM) requires two Higgs doublets from
supersymmetry\,\cite{carena}. Another particular mechanism, the spontaneous CP
breaking\,\cite{lee}, generally needs more scalars to be implemented.
Historically, the quest for alternative or additional CP violating sources was
the reason to consider simple extensions of the SM containing more than one
Higgs doublets, in particular, two and three Higgs
doublets\,\cite{lee,weinberg:scpv,WW}.

This work aims the study of the scalar potential of extensions of the SM with
$N$ Higgs doublets
(NHDMs)\;\cite{barrosoferreira:nhdm,lavoura:94,ccn:nhdm}. Such
models contain a reparametrization freedom\,\cite{ginzburg} induced by
$SU(N)_H$ transformations on the $N$ Higgs doublets which is physically
irrelevant because they are in the same representation of the gauge group, i.e.,
they possess the same quantum numbers. Such reparametrization transformations
are called horizontal transformations, acting on the horizontal space formed by
the $N$-Higgs doublets\,\cite{ccn:nhdm}. Hence, two different potentials defined
by two different sets of parameters but connected by some reparametrization
transformation are physically equivalent. Properties such as CP symmetry or
asymmetry is also independent of reparametrization which means any CP invariant
potential, even with complex parameters, can be connected to a potential where
all coefficients are real, i.e., manifestly CP symmetric\,\cite{GH}. Thus,
reparametrization invariant quantities, such as the Jarlskog
invariant\,\cite{jarlskog} in the SM, can be constructed to quantify CP
violation\,\cite{botella:94,branco:05,ccn:nhdm}.
In Ref.\,\onlinecite{ccn:nhdm}, we tried to solve the question: what are the
necessary and sufficient conditions for explicit and spontaneous CP violation
for a given NHDM potential? We could solve partially the explicit CP violation
conditions but the study of the different minima of the potential were not
considered.

Concerning general NHDM potentials we can pose two questions: (1) how to find
all the minima for a given potential specified by given parameters and (2) how
to parametrize all physically permissible or interesting NHDM potentials and
sweep all their parameter space. This work solves neither question (1) nor
question (2) completely, but some sufficient physical conditions can be
implemented and several consistency criteria can be formulated concerning
question (2) while question (1) can be solved in some classified cases.
Following the formalism adopted in Ref.\,\onlinecite{ccn:nhdm} to study CP
violation, and the extension for 2HDMs studied in
Ref.\,\onlinecite{ivanov:lorentz}, we will study the structure of NHDMs and the
properties of the different nontrivial minima. These different minima can be
first classified into two types: the usual neutral (N) minimum and the charge
breaking (CB) minimum. The former can be further classified into neutral normal
(NN) and CP breaking (CPB) minimum. With only one Higgs doublet, only the
neutral normal minimum is possible. With more than one doublet, emerges the
possibility of breaking also the electromagnetic symmetry
(CB)\,\cite{ferreira:2hdm:CBV,barrosoferreira:2hdm:noCBV} or the CP symmetry
(CPB)\,\cite{lee} spontaneously. Since the CP properties were already considered
in Ref.\,\onlinecite{ccn:nhdm}, we will concentrate on the differences between
charge breaking and neutral vacua.

A rich mathematical structure also emerges from the study of the NHDM
potential. We will see an underlying Minkowski structure will emerge,
analogously to the 2HDM potential\,\cite{ivanov:lorentz}. For example, it will
be shown that the variable space lies inside and on the future lightcone for an
appropriately chosen set of $N^2$ real variables.
The Minkowski structure, however, will not be sufficient to characterize
all the vacuum properties for $N>2$. Nevertheless, it is possible to consider
the Lorentz group $SO(1,N^2\!-\!1)$, containing the group $SU(N)_H$, as a
powerful parametrization tool. For example, a sufficient condition for bounded
below potentials can be formulated within this context.
Various properties of the vacuum, such as the distinction between charge
breaking and neutral vacuum, can be also formulated in a Minkowskian language.
Using a certain gauge choice, we will also see that the variable space can be
parametrized by two future lightlike vectors for which the cosine of the
angle between their spatial parts is the rational number $-(N\!-\!1)^{-1}$.

From the physical point of view, interesting predictive information can be
extracted for certain limits. For example, there are models preserving EM
symmetry which exhibits a degenerate mass spectrum for physical charged Higgs
bosons.

The outline is as follows: In Sec.\,\ref{sec:structure} we analyze the Minkowski
structure of the NHDM potentials and introduce some useful mathematical
definitions. The section is divided in the analysis of the variable space
(Sec.\,\ref{subsec:r.mu}) and the parameter space
(Sec.\,\ref{subsec:minkowski}). In Sec.\,\ref{sec:spoints} we analyze the
stationary points of the potential, introduce the physical charged Higgs
basis (Sec.\,\ref{subsec:PCH}) and analyze the properties of charge breaking
(Sec.\,\ref{subsec:CBV}) and neutral (Sec.\,\ref{subsec:NV}) vacua.
The conclusions are discussed in Sec.\,\eqref{sec:concl}.

%%%%%%%%%%%%%%%%%%%%%%%%%%%%%%%%%%%%%%%%%%%
\section{The structure of the NHDM potential}
\label{sec:structure}

In a previous work\,\cite{ccn:nhdm}, it was shown that a general gauge
invariant potential with $N\ge 2$ SM Higgs-doublets
$\Phi_a=(\phi_{a1},\phi_{a2})^{\tp}$, $a=1,\dots, N$, can be written solely in
terms of the real variables
\eq{
\label{A.mu}
\bbA^\mu\equiv \mn{\frac{1}{2}}\Phi^\dag_a (\lambda^\mu)_{ab}\Phi_b~,
~~\mu=0,1,\ldots,d,
}
where $\lambda^0=\sqrt{\frac{2}{N}}\,\id$ and $\{\lambda^i\}$ are the
$d=N^2-1$ hermitian generators of $SU(N)_H$ in the fundamental
representation, obeying the normalization
$\Tr[\lambda^\mu\lambda^\nu]=2\delta^{\mu\nu}$.
There is, nevertheless, a more appropriate normalization of the variable
$\mathbb{A}^0$ in Eq.\,\eqref{A.mu}, when $N>2$, which allows us to uncover a
Minkowski structure in the variable space of the NHDM potential, extending then
the 2HDM case\,\cite{ivanov:lorentz}.

Defining
\eq{
\label{r.mu}
r^\mu(\Phi)\equiv \Phi_a^\dag (T^\mu)_{ab}\Phi_b~,
~~\mu=0,1,\ldots,d,
}
where 
\eq{
T^\mu\equiv(\sqrt{\frac{\n\!-\!1}{2\n}}\id_{\mt{N}},\frac{1}{2}\lambda^i)\,,}
it is proved in the appendix \ref{proof:1} that 
\eq{
\label{r^2>=0}
r_\mu r^\mu=
\phi^*_{a1}\phi_{a1}\phi^*_{b2}\phi_{b2}
-|\phi^*_{a1}\phi_{a2}|^2
\ge 0\,,
}
assuming the usual Minkowski metric $g_{\mu\nu}=\diag(1,-\id_d)$, the definition
of the covariant vector $r_\mu\equiv g_{\mu\nu}r^\mu$ and the conventional sum
over repeated indices. Equation \eqref{r^2>=0} then restricts the space of the
variables $r^\mu$ to be inside and on the future lightcone 
\eq{
\lc\equiv\{x^\mu\in \mink\,|\,x^\mu x_\mu\ge 0, x_0>0\}\,,
}
in a Minkowski spacetime $\mink$.
We will see in sec.\,\ref{subsec:r.mu} that the variables $r^\mu=r^\mu(\Phi)$ in
Eq.\,\eqref{r.mu} do not cover the whole $\lc$ neither do they form a vector
subspace. It is important to stress that the quantity in Eq.\,\eqref{r^2>=0}
calculated for the vacuum expectation value signals a charge breaking
vacuum for nonzero values\,\cite{ivanov:lorentz}.

Using the Minkowski variables of Eq.\,\eqref{r.mu} we can write the most
general gauge invariant potential in the form
\eq{
\label{V:r.mu}
V(r)=M_\mu r^\mu+\mn{\frac{1}{2}}\Lambda_{\mu\nu}r^\mu r^\nu\,,
}
where $M^\mu$ is a general vector and $\Lambda^{\mu\nu}$ is a general
symmetric rank-2 tensor in Minkowski space.
The relation between the parameters $M$ and $\Lambda$ and the more
usual parameters $Y$ and $Z$, used to write the potential in the
form\,\cite{GH,endnote0}
\eq{
\label{V:Phi}
V(\Phi)=
Y_{ab}\Phi^\dag_{a}\Phi_b
+\mn{\frac{1}{2}}Z_{(ab)(cd)}(\Phi_{a}^\dag\Phi_b)^*
(\Phi_{c}^\dag\Phi_d) ~,
}
can be found in appendix \ref{ap:trans}.
The explicit parametrization for the 2HDM can be found in 
Ref.\,\onlinecite{ccn:nhdm}.

%%%%%%%%%%%%%%%%%%%%%%%%%%%%%%%%%%%%%%%%%%%5
\subsection{Variable space}
\label{subsec:r.mu}

The vector $r^\mu$ in Eq.\,\eqref{r.mu} defines a particular mapping of
$\{\Phi_a\}$ in $\mathbb{C}^N\otimes\mathbb{C}^2$ into
$\mink$. The former space can be parametrized by
$4(N\!-\!1)$ real parameters, with the $SU(2)_L\otimes U(1)_Y$ gauge freedom
already
taken into account, while the latter space requires $N^2=d+1$ parameters.
Since $N^2\ge 4(N\!-\!1)$ for $N\ge 2$, the mapping is obviously not surjective.
The image of such mapping defines therefore a space
\eq{
\mathcal{V}_{\Phi}\equiv\{x^\mu\in\lc|x^\mu=r^\mu(\Phi)\}\,,
}
contained in \lc. We will then analyze the properties of $\mathcal{V}_{\Phi}$
and seek a criterion to identify if a vector $x^\mu$ in \lc\ is also in
$\mathcal{V}_{\Phi}$.

Firstly, define the bijective mapping $f^\mu$ from the set of
hermitian complex $N\!\times N$ matrices, denoted by $\mathcal{M}_h(N,c)$,
into $\mink$:
\eq{
f^\mu(h)\equiv \Tr[T^\mu h]\,.
}
This mapping is invertible and therefore bijective, since, defining
\eq{
\tilde{x}\equiv 2x^\mu \tilde{T}_{\mu}\,,
}
where 
\eq{
\label{Ttilde}
\tilde{T}^\mu\equiv (\frac{T^0}{\n-1},-T^i)\,,
}
we identify
\eq{
h=\tilde{x}\,,
}
once the equality $f^{\mu}(h)=x^\mu$ holds.
Such identity can be easily verified by using the relation
\eq{
2\Tr[T^\mu \tilde{T}^\nu]=g^{\mu\nu}\,.
}

We can express the Minkowski inner product in $\mathcal{M}_h(N,c)$ by defining
a function $\Delta$ of an hermitian matrix $h$ as 
\eq{
\label{Delta}
\Delta(h)\equiv \mn{\frac{1}{2}}[(\Tr h)^2-\Tr(h^2)]\,.
}
It is easy to verify using the trace properties of $\tilde{T}^\mu$ that 
\eq{
\label{x^2.Delta}
x^\mu x_\mu=\Delta(\tilde{x})\,.
}
It is only for the particular case of $N=2$ that we have
$\Delta(\tilde{x})=\det\tilde{x}$, allowing the extension from $SU(2)$ to
$SL(2,c)$ that preserves the Minkowski metric and therefore can represent the
group of proper Lorentz transformations.

Now we can realize the definition in Eq.\,\eqref{r.mu} corresponds to the
$f^\mu$ mapping of a particular class of hermitian matrices. Defining vectors
$u$ and $w$ in $\mathbb{C}^N$ such that
\eqarr{
\label{u}
u_a &\equiv& \phi_{a1}\,, \\
\label{w}
w_a &\equiv& \phi_{a2}\,, 
}
we can see that
\eq{
r^\mu(\Phi)= f^\mu(uu^\dag+ww^\dag)
=f^\mu(uu^\dag)+f^\mu(ww^\dag)\,.
}
From Eq.\,\eqref{x^2.Delta} and the property $h^2=\Tr[h]\,h$ for $h=uu^\dag$, we
see $f^\mu(uu^\dag)$ and $f^\mu(ww^\dag)$  lie on the future lightcone.
Thus $r^\mu(\Phi)$ is a sum of two future lightlike vectors:
\eq{
\label{r.mu:xy}
r^\mu(\Phi)=x^\mu+y^\mu~, 
}
where $x^\mu x_\mu=0,~y^\mu y_\mu=0,~x^0,y^0>0$. Note that the splitting of
Eq.\,\eqref{r.mu:xy} into the sum of $x^\mu=f^\mu(uu^\dag)$ and
$y^\mu=f^\mu(ww^\dag)$ is not gauge invariant since $SU(2)_L$ gauge
transformations can mix $u$ with $w$.

Now we can state the \textbf{criterion}: 
\begin{quote}
a vector $x^\mu$ in \lc\ is also in $\mathcal{V}_\Phi$ if, and only
if, the corresponding matrix $\tilde{x}$ has rank two or less and its nonzero
eigenvalues are positive. A vector $x^\mu$ in $\mathcal{V}_\Phi$ is future
lightlike if, and only if, $\tilde{x}$ has rank one.
\end{quote}
The proof for necessity is trivial, since any matrix of the form $h=uu^\dag +
ww^\dag$ has rank two or less and its non-null eigenvalues are positive. The
converse can be proved by diagonalizing $\tilde{x}$. If $\tilde{x}$ has rank two
or less and its non-null eigenvalues are positive, it can be written in the form
\eq{
\label{x:diag}
\tilde{x}=\lambda_1^2 v_1v_1^\dag+\lambda_2^2 v_2v_2^\dag\,,
}
where $\lambda_i^2$ are the positive eigenvalues and $v_i$ their
respective normalized eigenvectors. 
With the identification $u=\lambda_1v_1$ and $w=\lambda_2v_2$ we see 
$x^\mu=f^\mu(uu^\dag+ww^\dag)$ is in $\mathcal{V}_\Phi$ and we complete our
proof. Setting $\lambda_2$ to zero and using Eq.\,\eqref{x^2.Delta}, we obtain
the rank one subcase.
One last remark concerns the ambiguity in associating $\tilde{x}$ with
$h=uu^\dag+ww^\dag$, since $u$ and $w$ need not to be orthogonal. 
However, the gauge freedom allows us to choose a particular representative of
$\Phi$ for which $u,w$ are orthogonal, i.e.,
\eq{
\label{u.w=0}
u^\dag w=0\,.
}
The proof is shown in appendix \ref{ap:gauge}. With the choice of
Eq.\,\eqref{u.w=0}, the mapping between $x^\mu$ in $\mathcal{V}_{\Phi}$ and
$h=uu^\dag+ww^\dag$ in $\mathcal{M}_h(N,c)$ is unambiguous, once an ordering
for the eigenvalues of $\tilde{x}$ is defined, hence $\Phi$ may also be
determined uniquely, except for rephasing transformations on $u,w$ which does
not alter the condition \eqref{u.w=0}. Thus $4(N\!-\!1)$ real parameters are
necessary to parametrize $u,w$ faithfully considering condition \eqref{u.w=0}
and the rephasing freedom for $u$ and $w$. Therefore, the same number of
parameters are necessary to parametrize $\mathcal{V}_\Phi$.
We will adopt the choice of Eq.\,\eqref{u.w=0} from this point on.
Since the sum of two rank two hermitian matrices can be equal or greater than
two, we also see $\mathcal{V}_{\Phi}$ does not form a vector subspace of
$\mink$. The exception happens for $N=2$ when they form a subspace
and $\mathcal{V}_\Phi=\lc$. 

An interesting feature arises with the adoption of Eq.\,\eqref{u.w=0}:
the cosine of the angle between the spatial parts of $x^\mu=f^\mu(uu^\dag)$ and
$y^\mu=f^\mu(ww^\dag)$ is a rational number. Such property can be seen by
\eq{
\label{r^2=2x.y}
r^\mu(\Phi)r_\mu(\Phi)=2x_\mu y^\mu=2x^0y^0(1-\cos\theta)\,,
}
where $\cos\theta=\displaystyle\frac{\mathbf{x\cdot y}}{|\bf x||\bf y|}$.
Equations \eqref{r^2>=0} and \eqref{x^2.Delta} imply
\eq{
\Delta(uu^\dag+ww^\dag)=|u|^2|w|^2=\frac{2\n}{\n-1}x^0y^0\,,
}
which yields the relation
\eq{
\label{cos:N}
\cos\theta=\frac{-1}{\n-1}\,.
}
The specific angles vary from $\theta=\pi$ ($N=2$) to $\theta \rightarrow
\pi/2^+$ ($N\rightarrow \infty$). In particular, for $N=2$, the vectors
$x^\mu$ and $y^\mu$ lie in opposite directions on the future lightcone.

%%%%%%%%%%%%%%%%%%%%%%%%%%%%%%%%%%%%%%%%%%%
\subsection{Parameter space}
\label{subsec:minkowski}

There are two advantages of parametrizing the potential in the form of
Eq.\,\eqref{V:r.mu} compared with the parametrization of Eq.\,\eqref{V:Phi}.
Firstly, we can consider any vector $M$ with $N^2$ components and any symmetric
tensor $\Lambda$ with $N^2\times N^2$ entries as parameters, restricted only by
physical requirements which will be further discussed, while the tensor
$Z_{(ab)(cd)}$ in Eq.\,\eqref{V:Phi} contains redundancies by
index exchange\,\cite{ccn:nhdm}. Therefore, we can adopt the parametrization of
Eq.\,\eqref{V:r.mu} as the starting point to analyze physical features such as
the requirement of bounded below potential or the possibility of having charge
breaking (CB) or CP breaking (CPB) vacua.

We have at our disposal $N^2(N^2\!+\!1)/2$ real parameters in $\Lambda$ and
$N^2$ real parameters in $M$. The number of physically significant parameters,
however, is fewer due to the reparametrization freedom which identifies all
potentials connected by horizontal transformations as physically equivalent. In
this context, the relevant horizontal group is $SU(N)_H$\,\cite{ccn:nhdm},
acting on the horizontal space spanned by the Higgs doublets.
The action of a horizontal transformation $U$ in the fundamental
representation $\mathbf{N}$ of $SU(N)_H$ can be written as
\eq{
\label{Phi:U}
\Phi_a \rightarrow U_{ab}\Phi_b\,.
}
While the quadratic variables $r^\mu$, transform leaving $r^0$ invariant and 
$r^i$ transforming according to the adjoint representation $\mathbf{d}$ of
$SU(N)_H$, in accordance to the branching $\bar{\mathbf{N}}\otimes \mathbf{N}=
\mathbf{d}\oplus {\bf 1}$.
Since adj$SU(N)_H$ can be obtained by exponentiation of the algebra spanned by
$i(T_j)_{kl}=f_{jkl}$ which is real and antisymmetric, adj$SU(N)_H$ forms a
subgroup of $SO(d)$.

Due to the $SU(N)_H$ reparametrization freedom, since the action of ${\rm
adj}SU(N)_H$ is effective on \lc, i.e., some orbits in \lc\ are not
trivial, the physically distinct potentials can be parametrized by only
$N^2+\frac{1}{2}N^2(N^2\!+\!1)-(N^2\!-\!1)=\frac{1}{2}N^2(N^2\!+\!1)+1$ real
parameters\,\cite{endnote01}. 
For $N=2$, such minimal number of parameters can be easily achieved by
diagonalizing the $3\times 3$ matrix $\Lambda_{ij}$, which gives $11$ parameters
needed to define $M$ (4), $\Lambda_{00}$ (1), $\Lambda_{0i}$ (3) and
$\Lambda_{ij}$ (3). When the potential exhibits CP invariance, such basis,
called canonical CP basis in Ref.\,\onlinecite{ccn:nhdm}, coincides with the
real basis\,\cite{GH} for which all coefficients in the potential are real. The
minimal parametrization for $N>2$ is not explicitly known\,\cite{ccn:nhdm}.

The second advantage of Eq.\,\eqref{V:r.mu} concerns the possibility of
extending $\mathrm{adj}SU(N)_H$ to $SO(d)$ and then to $SO(1,d)$ which is the
group of homogeneous proper Lorentz transformations in $\mink$.
The importance of such extension relies on the fact that $SO(1,d)$ leaves \lc\
invariant and acts transitively on it, i.e., any two vectors $x^\mu,y^\mu$ in
\lc\ can be connected by $SO(1,d)$.
If the parameter space generated by $r^\mu(\Phi)$ covered the whole \lc, we
could parametrize all physically inequivalent NHDM potentials by parametrizing
the cosets $SO(1,d)/{\rm adj}SU(N)_H$ acting on some fixed representative
classes of $\{M,\Lambda\}$.
For example, for $N=2$, all \lc\ can be covered by $r^\mu(\Phi)$ and all
physically bounded below potentials can be parametrized by parameters $M$ (4
parameters), $\Lambda=\diag(\Lambda_0,\Lambda_i)$ (4 parameters), with
$\Lambda_i>-\Lambda_0$, and a boost parameter $\vec{\xi}$ (3 parameters),
needed to generate the $\Lambda_{0i}$ components\,\cite{ivanov:lorentz}.
Boosts belong to $SO(1,3)/{\rm adj}SU(2)_H$ and, furthermore, specially for
$N=2$, they can be implemented over $\Phi$ with the extension of $SU(2)_H$ to
$SL(2,c)$.

Nevertheless, although the permissible variable space only covers
$\mathcal{V}_{\Phi}$, which is smaller than \lc\ when $N>2$, we can cover a
large class of physically acceptable potentials by considering all $r^\mu$ in
\lc\ and imposing the physical restrictions on the set $\{M,\Lambda\}$. The
physical restrictions to consider are (i) bounded below potential and (ii) the
existence of nontrivial extrema, $\aver{\Phi}\neq 0$.

We can impose the restriction (i) by requiring\,\cite{ivanov:lorentz}
\begin{itemize}
\item[P1:] $\Lambda$ is diagonalizable by $SO(1,d)$, i.e., there is a basis
where
\eq{
\Lambda_{\mu\nu}=\diag(\Lambda_0,\Lambda_i)\,,
}
\item[P2:] $\Lambda_{0}>0$ and $\Lambda_{i}>-\Lambda_{0}$.
\end{itemize}
The conditions P1 and P2 are necessary and sufficient to guarantee the quartic
part of the potential in Eq.\,\eqref{V:r.mu} to be positive definite for all
$r^\mu$ in \lc. Since the variable space does not cover the whole \lc\ but
only $\mathcal{V}_\Phi$, for $N>2$, the above conditions are only
\textbf{sufficient} to guarantee the positivity of the quartic part of the
potential. Obviously, the class of potentials with the quartic
part positive definite for all $r^\mu$ in $\mathcal{V}_\Phi$ is larger.
The proof of P1 and P2 follows analogously to the 2HDM case where the
group is $SO(1,3)$\,\cite{ivanov:lorentz}. The treatment of
general diagonalizable tensors in $SO(1,n-1)$ can be found in
Ref.\,\onlinecite{renardy}. 

The restriction (ii) of nontrivial extrema will be considered in the next
section where the properties of stationary points will be analyzed.
%% added 17.09.07
One can say, however, that to ensure the existence of nontrivial stationary
points ($\langle{\Phi}\rangle\neq 0$), it is necessary to have the quadratic
part of the potential acquiring negative values for some $\Phi$. The latter is
only possible when $Y$ in Eq. (7) has at least one negative eigenvalue.

%%%%%%%%%%%%%%%%%%%%%%%%%%%%%%%%%%%%%%%%%%%
\section{Stationary points}
\label{sec:spoints}

To find the stationary points we differentiate $V$ in Eq.\,\eqref{V:r.mu}:
\eq{
\label{dV.dphi}
\frac{\partial}{\partial \phi_{ai}^*}{V(\Phi)}=
\frac{\partial}{\partial r^\mu}{V(r)}
\frac{\partial r^\mu}{\partial \phi_{ai}^*}
=\mathbb{M}_{ab}\phi_{bi}\,,
}
where
\eqarr{
\label{bbM}
\mathbb{M}&\equiv & X_\mu T^\mu\,,\\
\label{X.mu}
X_\mu(r^\mu) &\equiv & M_\mu + \Lambda_{\mu\nu}r^\nu\,.
}
The stationary points $\aver{\Phi}$ correspond to the roots of
Eq.\,\eqref{dV.dphi}, i.e., solutions of
\eq{
\label{Phi:extrema}
\aver{(\mathbb{M}\otimes\id_2)\Phi}=0\,,
}
which requires 
\eq{
\label{det=0}
\det\aver{\mathbb{M}}=0
}
for nontrivial solutions $\aver{\Phi}\neq 0$. The brackets $\aver{~~}$ mean to
take expectation values on all fields $\Phi$, including on $\mathbb{M}$.

Rewriting Eq.\,\eqref{Phi:extrema} in terms of $u,w$ in Eqs.\,\eqref{u} and
\eqref{w}, we have
\eq{
\label{Mu.Mw}
\aver{\mathbb{M}u}=0~, ~~
\aver{\mathbb{M}w}=0\,.
}
If $\aver{u}$ and $\aver{w}$ are non-null and noncollinear, Eq.\,\eqref{Mu.Mw}
means that $\aver{\mathbb{M}}$ has two zero eigenvalues and $\aver{u},\aver{w}$
are the respective eigenvectors. From 
\eq{
\aver{\Phi^\dag\Phi}=\aver{u^\dag u}+\aver{w^\dag w}\,,
}
it is necessary that at least one of $\aver{u}$ or $\aver{w}$ be non-null to
have a nontrivial vacuum expectation value (VEV).
We can then classify charge breaking (CB) and neutral (N) stationary
points depending on
\begin{itemize}
\item cond.\,CB: $\aver{r^\mu r_\mu}>0$. Equivalently, both $\aver{u}$ and
$\aver{w}$ are non-null and noncollinear.
\item cond.\,N: $\aver{r^\mu r_\mu}=0$. Equivalently, either $\aver{u}$ or
$\aver{w}$ is null or they are collinear.
\end{itemize}

On the other hand, multiplying $\aver{\Phi}^\dag$ on the left of
Eq.\,\eqref{Phi:extrema} yields
\eq{
\label{X.r=0}
\aver{X_\mu r^\mu}=0\,.
}
For $\aver{r^\mu}$ timelike, any vector orthogonal, with respect to the
Minkowski metric, have to be spacelike\,\cite{SR:book}. For $\aver{r^\mu}$
lightlike only (a) lightlike collinear vectors and (b) spacelike vectors can be
orthogonal\,\cite{SR:book}. Then we can classify the solutions of
Eq.\,\eqref{X.r=0} into three types, when $\aver{r^\mu}\neq 0$ and in \lc:
\begin{itemize}
\item[(I)] Trivial solution with $\aver{X_\mu}=0$ and $\aver{\mathbb{M}}=0$; EM
symmetry can be broken or not\,\cite{endnote02}.

\item[(II)] Solution with $\aver{X_\mu X^\mu}=0$, $\aver{X_\mu}\neq 0$;
EM symmetry is always preserved and $\aver{X}^\mu=\alpha \aver{r^\mu}$
corresponding to case (a).

\item[(III)] Solution with $\aver{X_\mu X^\mu}<0$, $\aver{X_\mu}\neq 0$; EM
symmetry can be broken ($\aver{r^\mu r_\mu}>0$) or not ($\aver{r^\mu r_\mu}=0$).
\end{itemize}
Note that type (I) solutions also correspond to the stationary points of
$V(r^\mu)$ with respect to $r^\mu$.

Let us consider some special cases:
For $N=2$, for which the identity $\det \tilde{x}=\Delta(\tilde{x})$ is valid,
there are only solutions of type (I) and (II) since
Eq.\,\eqref{det=0} imply $\aver{X_\mu X^\mu}=0$. Furthermore, any charge
breaking solution is of type (I).
For $N=3$, the type (III) solution is present and because we need two null
eigenvalues for $\aver{\mathbb{M}}$, $\aver{X^\mu}$ must be in the
cone defined by $(N\!-\!1)^2X_0^2-\mathbf{X}^2=0$, i.e., $\aver{X^\mu}$ is
spacelike. The proof is shown in appendix \ref{ap:X^2=0}.

Now we can seek the explicit solutions. For type (I) solutions, an explicit
expression can be given,
\eq{
\label{sol:I}
\aver{r^\mu}= -(\Lambda^{-1})^\mu_{\;\nu}M^\nu\,.
}
Of course, $\aver{r^\mu}$ should be restricted to $\mathcal{V}_\Phi$ which only
happens when $-M^\mu$ is in the image of $\mathcal{V}_\Phi$ by
$\Lambda^{\mu}_{\;\nu}$\,\cite{ivanov:lorentz}. 
If $\Lambda$ is not invertible, it is necessary to take the inverse only over
the non-null space.

For type (II) solutions, $\aver{r^\mu}$ should satisfy
\eq{
\label{sol:II}
\aver{\Lambda_{\mu\nu}r^\nu-\alpha r_{\mu}}= -M_\mu\,,
}
where $\alpha$ is an unknown parameter which has to be determined from
Eq.\,\eqref{sol:II} and the constraint that $r^\mu$ should be in
$\mathcal{V}_{\Phi}$.
Obviously, there may be more than one of such solutions with different
$\alpha$, as it is for the $N=2$ case\,\cite{ivanov:lorentz}.

The type (III) solutions are not explicitly expressible and involves nonlinear
equations in Eq.\,\eqref{Mu.Mw}.

Let us analyze the general properties of the potential expanded around any
stationary point. The expansion is induced by the replacements
\eqarr{
\Phi &\rightarrow& \Phi+\aver{\Phi}\,,\\
r^\mu &\rightarrow& r^\mu+ \aver{r^\mu}+ s^\mu\,,
}
where
\eqarr{
\label{s.mu:1}
s^\mu&\equiv& \aver{\Phi}^\dag T^\mu \Phi+\Phi^\dag T^\mu\aver{\Phi}\,,\\
\label{s.mu:2}
&=&
f^\mu(u\aver{u}^\dag) + f^\mu(w\aver{w}^\dag) +h.c.
}
Thus,
\eq{
\label{V:134}
V(\Phi+\aver{\Phi})=V_0+V_2+V_3+V_4\,,
}
where
\eqarr{
V_0&=&V(\aver{r^\mu})\,,\\
\label{V:2}
V_2&=&\Phi^\dag \aver{\mathbb{M}} \Phi 
+ \mn{\frac{1}{2}}\Lambda_{\mu\nu}s^\mu s^\nu\,,\\
\label{V:3}
V_3&=&\Lambda_{\mu\nu}s^\mu r^\nu\,,\\
\label{V:4}
V_4&=&\mn{\frac{1}{2}}\Lambda_{\mu\nu}r^\mu r^\nu\,.
}
To guarantee the stationary point is a local minimum, it is necessary and
sufficient to have the mass matrix after spontaneous symmetry breaking (SSB),
extractable from Eq.\,\eqref{V:2}, to be positive semidefinite.
On the other hand, due to Eq.\,\eqref{X.r=0} and the positivity of $V_4$, we
have 
\eq{ 
V_0=\mn{\frac{1}{2}}M_\mu\aver{r^\mu}
=-\mn{\frac{1}{2}}\Lambda_{\mu\nu}\aver{r^\mu}\aver{r^\nu}<0\,. 
}
The last inequality means any nontrivial stationary point lies deeper than the
trivial extremum $\aver{\Phi}=0$.

%%%%%%%%%%%%%%%%%%%%%%%%%%%%%%%%%%%%%%%%%%%
\subsection{Physical Charged Higgs basis}
\label{subsec:PCH}

We can write the potential \eqref{V:134} in an explicit basis where the
physical degrees of freedom can be more easily extracted.
For such purpose we choose the physical charged Higgs (PCH)
basis\,\cite{endnote1} where
\eqarr{
\label{PCH:<>}
\aver{w}&=&
\begin{pmatrix}
0\cr \vdots \cr 0 \cr |\aver{w}|
\end{pmatrix}
=|\aver{w}|e_{\nn}~, ~~\cr
\aver{u}&=&
\begin{pmatrix}
0\cr \vdots \cr |\aver{u}|\cr 0
\end{pmatrix}
=|\aver{u}|e_{\mt{N\!-\!1}}~, ~~
}
where $e_i$, $i=1,\ldots,N$ defined by $(e_i)_j=\delta_{ij}$ are the canonical
basis vectors. The module $|\aver{w}|$ denotes the square root of $\aver{w^\dag
w}=\aver{w}^\dag\aver{w}$.
Such choice is always allowed by the $SU(N)_H$ reparametrization
freedom, once the condition \eqref{u.w=0} is met. Although there is an
additional $SU(N\!-\!1)$ or $SU(N\!-\!2)$ reparametrization freedom in the
subspace orthogonal to $\aver{w}\neq 0$ or $\aver{w},\aver{u}\neq 0$, which need
to be fixed to specify the PCH basis. Conventionally, we choose $\aver{w}$ to be
always non-null from the requirement of nontrivial vacuum. Therefore,
$\aver{u}\neq 0$ or $\aver{u}=0$ correspond respectively to the charge breaking
vacuum (CBV) and the neutral vacuum (NV) solutions. 

In the PCH basis 
\eqarr{
\label{ww:n.mu}
f^\mu(\aver{ww^\dag})&=&\aver{w^\dag w}\mn{\sqrt{\frac{N\!-1}{2N}}}n^\mu\,,\\
\label{uu:n'.mu}
f^\mu(\aver{uu^\dag})&=&\aver{u^\dag u}\mn{\sqrt{\frac{N\!-1}{2N}}}n'{}^\mu\,,
}
where $n^\mu$ and $n^{\pr\mu}$ have non-null components
\eqarr{
\label{n.mu}
(n^0,n^{\mt{N\!-\!2}},n^{\mt{N\!-\!1}})&=& (1,0,-1)\,,\\
\label{n'.mu}
(n'{}^{0},n'{}^{\mt{N\!-\!2}},n'{}^{\mt{N\!-\!1}})&=& 
(1,-\mn{\frac{\sqrt{N(N\!-2)}}{N\!-1}},\mn{\frac{1}{N\!-1}})\,,
}
given 
%$r\equiv N\!-\!1$ (not to be confused with $r^\mu$) and 
the ordering of $\mu$ following
\eq{
\label{T.mu:list}
\{T^0,h_a,\mathcal{S}_{ab},\mathcal{A}_{ab}\}\,,
}
with $a=1,\ldots,N\!-\!1$, $b=1,\ldots,N$, and $a<b$, denoting the non-null
entries of $2(\mathcal{S}_{ab})_{ab}=2(\mathcal{S}_{ab})_{ba}=1$ and
$2(\mathcal{A}_{ab})_{ab}=-2(\mathcal{A}_{ab})_{ba}=-i$\,\cite{ccn:nhdm}, which
are the combination of ladder operators analogous to $\sigma^1$ and $\sigma^2$
for $SU(2)$. The matrices $h_a$ form the Cartan subalgebra which can be chosen
diagonal. Notice that Eqs.\,\eqref{n.mu} and \eqref{n'.mu} satisfy
Eq.\,\eqref{cos:N}.

From Eq.\,\eqref{Mu.Mw}, we have for $\aver{w}\neq 0$
\eq{
\label{MiN=0}
\aver{\mathbb{M}_{aN}}=\aver{\mathbb{M}_{Na}}=0\,,
}
for all $a=1,\ldots,N$. In addition, if $\aver{u}\neq 0$ (CBV), we have
\eq{
\label{Mir=0}
\aver{\mathbb{M}_{a,{\mt{N\!-\!1}}}}=\aver{\mathbb{M}_{{\mt{N\!-\!1}},a}}=0\,,
}
reducing the non-null matrix to its upper-left $(N\!-\!1)\times (N\!-\!1)$
($\aver{u}=0$) or $(N\!-\!2)\times(N\!-\!2)$ ($\aver{u}\neq 0$) submatrix.
In both cases we can use the remaining reparametrization freedom to choose
$\aver{\mathbb{M}}$ diagonal
\eq{
\label{bbM:diag}
\aver{\mathbb{M}}=
\left\{
\begin{array}{lll}
\diag(m^2_a,0,0), &a=1,\ldots,N\!-\!2 &\text{ for } \aver{u}\neq 0\,,\cr
\diag(m^2_a,0), &a=1,\ldots,N\!-\!1 &\text{ for } \aver{u}= 0\,.
\end{array}
\right.
}
This form can be always achieved because the remaining $SU(N\!-\!1)$ or
$SU(N\!-\!2)$ reparametrization freedom leaves $\aver{r^\mu}$ invariant.
Equation \eqref{bbM:diag} defines the PCH basis uniquely if the eigenvalues
$m^2_a$ are ordered, assuming they are not degenerate. 

The null eigenvalues of Eq.\,\eqref{bbM:diag} correspond to the Goldstone modes
for the combination of fields not present in $s^\mu(\Phi)$ in Eq.\,\eqref{V:2}.
The four massless Goldstone modes are
\eq{
\sqrt{2}\im(w_{\nn}),~~\sqrt{2}\im(u_{\mt{N\!-\!1}})
}
and the fields $R,I$ proportional, by real normalization constants, to
\eqarr{
\label{R}
R&\propto & |\aver{u}|\re(w_{\mt{N\!-\!1}})-|\aver{w}|\re(u_{\nn})\,,\\
\label{I}
I&\propto & |\aver{u}|\im(w_{\mt{N\!-\!1}})+|\aver{w}|\im(u_{\nn})\,.
}
To find the Goldstone modes for the neutral vacuum solution it is sufficient to
set $\aver{u}=0$ in the equations above and disconsider Eq.\,\eqref{Mir=0} which
make $\sqrt{2}\im(u_{\mt{N\!-\!1}})$ also massive. The explicit form of $s^\mu(\Phi)$ in
this basis is shown in appendix \ref{ap:s.mu}.

%%%%%%%%%%%%%%%%%%%%%%%%%%%%%%%%%%%%%%%%%%%
\subsection{Charge breaking vacuum}
\label{subsec:CBV}

A vacuum expectation value $\aver{\Phi}$ breaking EM symmetry
(CBV)\,\cite{ferreira:2hdm:CBV}, is characterized by cond.\,CB stated in
Sec.\,\ref{sec:spoints}. They can be of type (I) or (III).
To assure two zero eigenvalues we must have
\eq{
\label{M:CBV}
\det\aver{\mathbb{M}}=(-1)^{\mt{N\!-\!1}}\gamma_{\nn}(\aver{\mathbb{M}})=0\,, ~~
\gamma_{\mt{N\!-\!1}}(\aver{\mathbb{M}})=0\,.
}
The explicit forms of the matricial functions $\gamma_k$ are unimportant here,
except that knowing the traces $\Tr[\aver{\mathbb{M}}^j]$ from $j=1,\ldots,k$
determines $\gamma_k$ uniquely. The explicit form can be found in
Eq.\,\eqref{gamma.k}.
Equation \eqref{M:CBV} defines two equations for $\aver{r^\mu}$ in addition to
the restriction that $\aver{r^\mu}$ belongs to $\mathcal{V}_\Phi$. Then, 
the possible vectors $\aver{u}$ and $\aver{w}$ extracted from the possible
$\aver{r^\mu}$, through the procedure in Eq.\,\eqref{x:diag}, should be
the eigenvectors of $\aver{\mathbb{M}}$ with eigenvalue zero.

Some conditions, however, can be extracted in the PCH basis.
From $\aver{w^\dag\mathbb{M}w}=0$ and $\aver{u ^\dag\mathbb{M}u}=0$, we have,
respectively,
\eqarr{
\label{X:0r}
\aver{X_0}&=&\aver{X_{\mt{N\!-\!1}}}\,,\\
\label{X:0rr-1}
\sqrt{N(N\!-\!2)}\aver{X_{\mt{N\!-\!2}}}&=&
(N\!-\!1)\aver{X_0}+\aver{X_{\mt{N\!-\!1}}}\cr
&=& N\aver{X_0}\,.
}
Then,
\eq{
\label{X^2:CB}
-\aver{X_\mu X^\mu}\ge
\aver{X_{\mt{N\!-\!1}}^2+X_{\mt{N\!-\!2}}^2-X_0^2}
=\frac{N}{N\!-\!2}\aver{X_0^2}\,,
}
confirming, for $N>2$, that all charge breaking solutions are of type (III)
unless $\aver{X_i}=0$, which implies a type (I) solution.

For type (I) solutions, one can see from Eq.\,\eqref{V:2} that the masses of
all scalars will depend only on $\Lambda$ which has to be positive definite in
the basis defined by the non-Goldstone fields; such condition assures the
stationary point is a local minimum.
In the PCH basis we can extract the mass matrix from the field combinations
$s^\mu(\Phi)$ in appendix \ref{ap:s.mu}. 
The only non-null combinations are $s^\mu(\Phi)$ with 
\eq{
\label{PCH:s.mu!=0}
T^\mu=T^0,h_{\mt{N\!-\!2}},h_{\mt{N\!-\!1}},
\mathcal{S}_{aN},\mathcal{S}_{b{\mt{N\!-\!1}}},\mathcal{A}_{aN},
\mathcal{A}_{b{\mt{N\!-\!1}}}\,,
}
for $a=1,\ldots,\ms{N\!-\!1}$ and $b=1,\ldots,\ms{N\!-\!2}$.
These field combinations can be considered as independent except for
\eq{
\label{sr-1:s0sr}
s^{\mt{N\!-\!2}}= -\sqrt{\frac{\n\!-\!2}{\n}}(s^{0}+s^{\mt{N\!-\!1}})\,.
}
The mass matrix $(M^2_{CB})_{ab}$ can then be extracted from $\Lambda_{\mu\nu}$
eliminating all components $\mu,\nu$ not contained in Eq.\,\eqref{PCH:s.mu!=0}
and eliminating the component $\mu=N-2$ or $\nu=N-2$ using
Eq.\,\eqref{sr-1:s0sr}. The resulting matrix, which is $4(N-1)$ dimensional
[$1+1+2(N-1)+2(N-2)$], should be positive definite. For $N=2$, $(M^2_{CB})_{ab}$ is
four dimensional and is $\Lambda_{\mu\nu}$ itself, identifying
$a,b=\mu+1,\nu+1=1,2,3,4$, except for normalization factors for
$s^\mu(\Phi)$\,\cite{ivanov:lorentz}.

For type (III) solutions, in addition to the second term of Eq.\,\eqref{V:2},
which is the same as for type (I) solutions, we have to add the first term
given by 
\eq{
\sum_{a=1}^{\mt{N\!-\!2}}m^2_a(|u_a|^2+|w_a|^2)\,,
}
using Eq.\,\eqref{bbM:diag}.
Notice that the coefficients of $\Lambda_{\mu\nu}$, not present in
Eq.\,\eqref{V:2}, do not contribute to the masses but only to the trilinear and
quartic interactions in Eqs.\,\eqref{V:3} and \eqref{V:4}.

%%%%%%%%%%%%%%%%%%%%%%%%%%%%%%%%%%%%%%%%%%%
\subsection{Neutral vacuum}
\label{subsec:NV}

A neutral vacuum (NV) is characterized by cond.\,N stated in
Sec.\,\ref{sec:spoints}. These solutions have $\aver{r^\mu(\Phi)}$
lightlike, $\aver{w}\neq 0$ but $\aver{u}=0$ and they can be of types (I), (II)
or (III). We can set $\aver{u}=0$ in all previous calculations where charge
breaking were assumed. We can promptly see that $s^\mu$ in Eq.\,\eqref{s.mu:2}
does not depend on $u_a$. Hence, from Eq.\,\eqref{V:2} we conclude that
$\aver{\mathbb{M}}$ is the mass matrix for the charged Higgs bosons, i.e.,
the matrix whose eigenvalues are the squared masses of the charged Higgs bosons,
combinations of $u_a$. The single null eigenvalue corresponds to the charged
Goldstone. 
%modified after prd submission; not contained in the referee response
This conclusion can be also reached by taking the matrix of second
derivatives of $V$ with respect to $\phi^*_{ai}$ and $\phi_{bj}$, and take
the VEV for $i=j=1$.
%%%
On the other hand, the mass matrix for neutral Higgs bosons, combinations of
$w_a$, depends explicitly on $\Lambda$ in addition to the contribution of
$\aver{\mathbb{M}}$. In the PCH basis, the three Goldstone modes are the neutral
$\sqrt{2}\im w_N$ and charged $u_N$. The SM Higgs is $\sqrt{2}\re w_{N}$.

Let us analyze type (III) solutions for which the following
proposition can be proved.
\begin{quote}
\noindent
\textbf{Proposition 1}:
For all $N\ge 3$, any type (III) solution which preserves EM symmetry must have
$\aver{X^\mu}$ in the region defined by 
\eq{
LC_N=\{x^\mu\in \mink|\,(N\!-\!1)^2x_0^2-\mathbf{x}^2\ge 0 
\text{~and~} x_\mu x^\mu <0\}
\,.
}
\end{quote}
This conditions is not Lorentz invariant but $SU(N)_H$ invariant. Such
proposition means neutral type (III) solutions can not have arbitrarily
spacelike $\aver{X^\mu}$. The proof is shown in appendix \ref{ap:X^2=0}.

The type (II) solutions are the most predictive ones for we have
$\aver{X^\mu}=\alpha\aver{r^\mu}$, $\alpha>0$.
From 
\eq{
2x_\mu T^\mu=\frac{\n}{\n-1}T^0x_0 - \tilde{x}\,,
}
for any $x^\mu$ in $\mink$, we can conclude that 
\eqarr{
\label{M:NV}
\aver{\mathbb{M}}&=&\alpha\aver{r_\mu}T^\mu\cr
&=&
\frac{\alpha \aver{\Phi^\dag \Phi}}{2}
[\id - \frac{\aver{ww^\dag}}{\aver{w^\dag w}}]
\,,
}
where $\aver{w^\dag w}=\aver{\Phi^\dag\Phi}=v^2/2$ and $v=246 \rm GeV$ is the
electroweak symmetry breaking scale, considering the basis where $\aver{u}=0$.
Obviously, $\aver{w}$ is an eigenvector of $\aver{\mathbb{M}}$ with eigenvalue
zero. Notice that Eq.\,\eqref{M:NV} implies  $\aver{\mathbb{M}}$ satisfies the
matricial equation
\eq{
\aver{\mathbb{M}}^2=\frac{\alpha}{4} v^2\aver{\mathbb{M}}\,.
}

With the simple structure of Eq.\,\eqref{M:NV}, a remarkable result can be
proved: \textbf{all charged physical Higgs bosons have the same mass}. Such
result can be more easily seen in the PCH basis where Eq.\,\eqref{PCH:<>} is
valid. Then, from Eq.\,\eqref{M:NV}, the physical charged Higgs bosons are the
fields $u_i$, with $i=1,\ldots,\ms{N\!-\!1}$, and they all have mass squared
\eq{
m^2_{H^+}=\frac{\alpha}{4}v^2\,.
}
Although the exact value of $\alpha$ should be a complicated function of the
parameters $M,\Lambda$ derived from Eq.\,\eqref{sol:II}, the degenerate mass
spectrum is a testable prediction.

The mass matrix for the neutral fields can be also straightforwardly constructed
from $\aver{\mathbb{M}}$ and $\Lambda$ using Eq.\,\eqref{V:2} but usually
nondegenerate because of the contribution of $\Lambda$. The procedure of
construction, in the PCH basis, is analogous to the one in
Sec.\,\eqref{subsec:CBV} but the non-null components of $s^\mu$, instead of
Eq.\,\eqref{PCH:s.mu!=0}, correspond to
\eq{
T^\mu=T^0,h_{\mt{N\!-\!1}},\mathcal{S}_{aN},\mathcal{A}_{aN}\,,
}
$a=1,\ldots,N-1$, with the non-null $s^\mu$ all functionally independent
and depending solely on $w_a$. The procedure is the same for type (III)
solutions.

Comparing neutral type (II) solutions with neutral type (III) solutions, we see
$-\aver{X^\mu X_\mu}\ge 0$ is a measure of how degenerate are the masses of the
physical charged bosons. Knowing the mass matrix $\aver{\mathbb{M}}$, 
we can recover $\aver{X_\mu}$ from
\eq{
\aver{X_\mu}=2\Tr[\tilde{T}_\mu \aver{\mathbb{M}}]\,.
}

The properties of neutral type (I) solutions can be analyzed setting
$\alpha\rightarrow 0$ in the type (II) solutions. We can conclude
that all charged Higgs bosons are massless. Therefore, there are $N\!-\!1$
charged pseudo Goldstone bosons and one genuine charged Goldstone contributing
to the Higgs mechanism.

%%%%%%%%%%%%%%%%%%%%%%%%%%%%%%%%%%%%%%%%%%%
\section{Conclusions}
\label{sec:concl}

The study of the NHDM potentials performed here reveals a very rich underlying
structure. In terms of the set of variables defined in Eq.\,\eqref{r.mu}, 
the variable space is limited to a subregion contained inside and on the future
lightcone \lc\ of a $1+d=N^2$ dimensional Minkowski space. Furthermore,
imposing the gauge condition \eqref{u.w=0}, the variable space can be
parametrized by two lightlike vectors whose spatial parts form an angle for
which the cosine is $-(N\!-\!1)^{-1}$.
The Minkowski structure also enabled us to find a sufficient, yet very
general, criterion to require a bounded below potential. The Lorentz group can
be also used as a powerful parametrization tool using the cosets
$SO(1,d)/\mathrm{adj}SU(N)_H$ to avoid reparametrization redundancies.
Charge breaking vacuum and neutral vacuum can be distinguished by calculating
the Minkowski length of $r^\mu(\Phi)$ for VEVs. The stationary points can be
classified according to the Minkowski length of $\aver{X_\mu}$, in
Eq.\,\eqref{X.mu}, into types (I), (II) and (III). 

The Minkowski structure would also help to seek the type (II) minima. The method
of caustics presented in Ref.\,\onlinecite{ivanov:lorentz} may be generalized to
count the number of type (II) solutions for $r^\mu$ restricted to \lc. The
restriction to $\mathcal{V}_\Phi$, however, would need more mathematical tools.
For example, the proper parametrization of $SO(1,d)/\mathrm{adj}SU(N)_H$ would
be very important to the complete study of the NHDM potential minima.

The knowledge of the matrix $\aver{\mathbb{M}}$ (or $\aver{X_\mu}$) and
$\Lambda$ is sufficient to construct the mass matrix for all the scalars. In
particular, when EM symmetry is not broken, $\aver{\mathbb{M}}$ is itself the
mass matrix of the charged Higgs bosons while the mass matrix of neutral bosons
also requires the information of $\Lambda$. In view of the privileged
information contained in $\aver{\mathbb{M}}$, one can try to parametrize any
physical NHDM potential by attributing to $\aver{\mathbb{M}}$ a general $N\times
N$ hermitian matrix (positive semidefinite if NV) with one (NV) or two (CBV)
null eigenvalues and attributing to $\Lambda$ a general $N^2\times N^2$ real
symmetric matrix which keeps $V_4$ of Eq.\,\eqref{V:4} positive definite. The
quadratic coefficient before SSB, $Y=M_\mu T^\mu$, can be obtained from 
\eq{
\label{Y:M.r}
Y=\aver{\mathbb{M}}-\Lambda_{\mu\nu}\aver{r^\mu}T^\nu\,,
}
where
\eq{
\label{r.mu:M}
\aver{r^\mu}=\alpha_1f^\mu(\aver{v_1v_1^{\dag}})+
\alpha_2f^\mu(\aver{v_2v_2^{\dag}})\,,
}
with $\alpha_1,\alpha_2$ nonnegative and $v_1,v_2$ orthonormal eigenvectors of
$\aver{\mathbb{M}}$ with eigenvalue zero. The parameters $\alpha_1,\alpha_2$
should be constrained by $\alpha_1+\alpha_2=\aver{\Phi^\dag\Phi}=v^2/2$. This
parametrization is not minimal but it assures that the stationary point
\eqref{r.mu:M} is a local minimum and has the advantage that some physical
parameters, such as the masses of the charged Higgs bosons, can be chosen as
parameters. On the other hand, nothing prevents the potential,
defined with general $\Lambda$ and $Y$, as in Eq.\,\eqref{Y:M.r}, to have a
minimum $\aver{r'^\mu}$ that lies deeper than the original $\aver{r^\mu}$, in
Eq.\,\eqref{r.mu:M}, used for parametrization. Such possibility limits the
potentialities of this parametrization fixed by $\{\aver{\mathbb{M}},\Lambda\}$
since the original minimum must be checked if it is the absolute minimum.
In the 2HDM, for example, potentials with two neutral vacua lying in
different depths can be constructed\,\cite{barroso:2hdm:2NV}.

For parametrization purposes, the form of Eq.\,\eqref{V:r.mu} is also very
advantageous since it avoids the redundancies contained in $Z_{(ab)(cd)}$ when
written in the form of Eq.\,\eqref{V:Phi}. Other several quantities can guide,
for instance, numerical studies to distinguish charge breaking vacua from
neutral vacua or local minima from saddle points. To identify the absolute
minimum, however, is still a difficult question.

The interesting case of mass degenerate charged Higgs bosons, the type (II)
vacuum, may have testable phenomenological implications. Because of the same
mass we could have an enhancement of production of physical charged Higgs bosons
for large $N$. However, even in this case, because some parameters in $\Lambda$
can be functionally free in the trilinear and quartic interactions, the
predictions for its width can be very difficult and variable. Usually, as
expected, as $N$ grows, we rapidly lose predictability unless we impose some
symmetries or approximations. The mass degeneracy is then a very predictive
result for certain NHDMs.

Even without the knowledge of an explicit minimum of the potential, writing the
theory in the PCH basis presents various advantages. The two main advantages
are the easily extractable physical informations and the minimality of
parametrization. For example, the VEVs in the PCH basis depend only on two real
nonnegative parameters, $|\aver{u}|$ and $|\aver{w}|$, a smaller number than the
four real parameters needed in the basis shown in
Ref.\,\onlinecite{barrosoferreira:nhdm}. Obviously, since the VEVs are real, the
CP properties of the vacuum should be encoded in the parameters $M'$ and
$\Lambda'$ transformed by $SU(N)_H$ in the PCH basis. Thus, if the original
parameters $M$ and $\Lambda$ are invariant by the canonical CP
reflections\,\cite{ccn:nhdm}, only the real subgroup $SO(N)_H$ should connect
the original basis to the PCH basis, besides rephasing transformations.

In conclusion, the results presented here uncover a rich structure contained
in the NHDM potential and illuminates the properties of the possible vacua. A
complete study of certain specific models should be guided by more
restrictive ingredients and interesting phenomenology. The study performed here,
however, is sufficiently general to cover a large class of physically possible
NHDMs.

%%%%%%%%%%%%%%%%%%%%%%%%%%%%%%%%%%%%%%%%%%%
\appendix
\section{Proof of Eq.\,$\text{\normalsize\eqref{r^2>=0}}$}
\label{proof:1}

Firstly, we recall the completeness formulas for $SU(N)_H$ and
$SU(2)_L$ respectively\,\cite{ccn:fierz}
\eqarr{
\label{compl:SUN}
\mn{\frac{1}{2}}(\lambda^\mu)_{ab}(\lambda^\mu)_{cd}&=&
\delta_{ad}\delta_{cb}\,,\\
\mn{\frac{1}{2}}(\sigma^\mu)_{ij}(\sigma^\mu)_{kl}&=&
\delta_{il}\delta_{kj}\,.
}
Then, the combination of both relations yields
\eq{
\label{lambda.sigma}
\Phi^\dag(\lambda^\mu\otimes\id_2)\Phi
\Phi^\dag(\lambda^\mu\otimes\id_2)\Phi
=
\Phi^\dag(\id_N\otimes\sigma^\mu)\Phi
\Phi^\dag(\id_N\otimes\sigma^\mu)\Phi
\,,
}
where the lefthand side of the equation is a shorthand for 
$\phi^*_{ai}(\lambda^\mu)_{ab}\phi_{bi}
\phi^*_{ck}(\lambda^\mu)_{ab}\phi_{dk}$, the indices $a,b,c,d=1,\ldots, N$
label the doublets (horizontal space) and $i,k=1,2$ label each field in the
doublet [representation space for $SU(2)_L$].
We can explicitly verify the relation
\eq{
\label{sigma}
\sigma\equiv
\phi^*_{a1}\phi_{a1}\phi^*_{b2}\phi_{b2}-|\phi^*_{a1}\phi_{a2}|^2
=
\mn{\frac{1}{4}}[(\Phi^\dag\Phi)^2-(\Phi^\dag\id_{N}\otimes
\vec{\sigma}\Phi)^2]
\,,
}
which is always nonnegative due to Schwartz inequality.
Finally, substituting $(\Phi^\dag\id_N\otimes \vec{\sigma}\Phi)^2$ of
Eq.\,\eqref{sigma} into Eq.\,\eqref{lambda.sigma} yields
\eq{
\frac{\n-1}{2\n}(\Phi^\dag\Phi)^2-
(\mn{\frac{1}{2}}\Phi^\dag\vec{\lambda}\otimes\id_2\Phi)^2=
\sigma\ge 0
\,,
}
which is the desired relation.
%%%%%%%%%%%%%%%%%%%%%%%%%%%%%%%%%%%%%%%%%%%
\section{Translation rules}
\label{ap:trans}

The relation between the parameters $Y,Z$ of 
Eq.\,\eqref{V:Phi} and the parameters $M,\Lambda$ of Eq.\,\eqref{V:r.mu} 
is\,\cite{ccn:nhdm}
\eqarr{
M_\mu&=& 2\Tr[Y\tilde{T}_\mu]\,,\\
\Lambda_{\mu\nu}&=& 4(\tilde{T}_\mu)_{ab}Z_{(ab)(cd)}(\tilde{T}_\mu)^*_{cd}\,,
}
where $\tilde{T}^\mu$ is defined in Eq.\,\eqref{Ttilde}.
The relations above can be obtained from the inverse of Eq.\,\eqref{r.mu},
\eq{
\Phi_a^\dag\Phi_b=2(\tilde{T}_\mu)_{ba}r^\mu(\Phi)=2(\tilde{T}_\mu r^\mu)^*_{ab}\,,
}
derived from the completeness relation \eqref{compl:SUN} written in terms of $T^\mu$ 
and $\tilde{T}^\mu$, i.e.,
\eq{
2(T^\mu)_{ab}(\tilde{T}_\mu)_{cd}=\delta_{ad}\delta_{cb}\,.
}

%%%%%%%%%%%%%%%%%%%%%%%%%%%%%%%%%%%%%%%%%%%
\section{Gauge choice for $\Phi$}
\label{ap:gauge}

We will prove here we can always perform a gauge transformation $SU(2)_L\otimes
U(1)_Y$ on $\Phi=u\otimes e_1+w\otimes e_2$ that renders $u$ and $w$
orthogonal, i.e., Eq.\,\eqref{u.w=0} is satisfied.

Firstly, we recall a gauge transformation $U$ acts equally on all the doublets
as
\eq{
\Phi_a \rightarrow U\Phi_a\,.
}
We know some gauge transformations in $SU(2)_L$ mix the vectors
$u_a=\phi_{a1}$ and $w_a=\phi_{a2}$, which induces complicated transformations
on $f^\mu(uu^\dag)$ and $f^\mu(ww^\dag)$. We know, however, the combinations
\eq{
\label{z.A}
z_A=
\begin{pmatrix}u^\dag & w^\dag \end{pmatrix}
\tau_A
\begin{pmatrix}u \cr w \end{pmatrix}
\,,~~~A=1,2,3\,,
}
transform as vectors in $\mathbb{R}^3$ by ordinary rotations, where $\tau_A$ are
Pauli matrices, generators of $SU(2)_L$.
Therefore, we can always rotate the vector in Eq.\,\eqref{z.A} to its third
component. Requiring $z_1=2\re(u^\dag w)=0$ and $z_2=2\im(u^\dag w)=0$ implies
$u^\dag w=0$.

%%%%%%%%%%%%%%%%%%%%%%%%%%%%%%%%%%%%%%%%%%%
\section{Characteristic equation for matrices}
\label{ap:X^2=0}

Firstly, comparing
\eq{
\Delta(2x_\mu T^\mu)= (N\!-\!1)^2x_0^2-\mathbf{x}^2
}
with $\Delta(2x_\mu \tilde{T}^\mu)=x_\mu x^\mu$, we obtain
\eq{
x_\mu x^\mu=\Delta(2x_\mu T^\mu) - N(N\!-\!2)x_0^2\,.
}
Now we can equate $x_\mu=\aver{X_\mu}$ of Eq.\,\eqref{X.mu} and require
semidefinite positiveness for $\aver{\mathbb{M}}$, Eq.\,\eqref{bbM}, i.e., all
eigenvalues are nonnegative, since $\aver{\mathbb{M}}$ corresponds to the mass
matrix of the charged Higgs bosons when EM symmetry is preserved. Then, the
semidefinite positivity of $\aver{\mathbb{M}}$ implies
$\Delta(\aver{\mathbb{M}})\ge 0$, hence
\eq{
\label{LC.N}
\aver{X_\mu X^\mu}\ge -N(N\!-\!2)\aver{X_0}^2\,,
}
which is equivalent to state that $\aver{X^\mu}$ is in $LC_N$, when
$\aver{X}^\mu$ spacelike.
The equality holds for $N=3$, for charge breaking solutions, as will be proved
in the following.

For any square matrix $A$, the characteristic equation can be
written as\,\cite{ccn:nhdm} 
\eq{
\det(A-\lambda \id)=(-1)^n
[\lambda^n-\sum_{k=1}^{n}\gamma_k(A)\lambda^{n-k}]
~,
}
where
\eq{
\label{gamma.k}
\gamma_k(A)=\frac{1}{k}\Tr[A^k-\sum_{j=1}^{k-1}\gamma_j(A)A^{k-j}]
~.
}
In particular, the function $\Delta$ defined in Eq.\,\eqref{Delta} is related
to $\gamma_2$ by
\eq{
\Delta(A)=-\gamma_2(A)\,.
}

For $3\times 3$ matrices we have then
\eq{
\det(A-\lambda \id)=(-1)
[\lambda^3-\gamma_1(A)\lambda^{2}-\gamma_2(A)\lambda-\gamma_3(A)].
}
To have two null eigenvalues we must have $\gamma_2(A)=\gamma_3(A)=0$.

%%%%%%%%%%%%%%%%%%%%%%%%%%%%%%%%%%%%%%%%%%%
\section{\normalsize $s^\mu$ in the physical charged Higgs basis}
\label{ap:s.mu}

The combination of fields in $s^\mu(\Phi)$, Eq.\,\eqref{s.mu:1}, determines the
non-Goldstone modes. In the physical charged Higgs (PCH) basis, they can be
written explicitly. The list of all $T^\mu$ was shown in Eq.\,\eqref{T.mu:list},
as well as the explicit representation for $\mathcal{S}_{ab}$ and
$\mathcal{A}_{ab}$. For the Cartan subalgebra formed by $h_a$ we can
adopt\,\cite{ccn:nhdm}
\eq{
h_a=\mn{\frac{1}{\sqrt{2\ms{a(a+1)}}}}\diag(\id_a,-a,0,\ldots,0),~~
a=1,\ldots,N\!-\!1\,.
}
A more detailed description of the parametrization of the $SU(N)$ algebra in the
fundamental representation can be found in Ref.\,\onlinecite{ccn:nhdm}.
In the following, we list only the non-null components of $s^\mu$, according to
$T^\mu$.
\begin{itemize}
\item $T^0$:
\eq{
\label{s:0}
s^0=\mn{\sqrt{\frac{2(N\!-1)}{N}}}
\Big[|\aver{w}|\re(w_N)+|\aver{u}|\re(u_{\mt{N\!-\!1}})\Big]\,.
}

\item $T^i=h_a$, $a=1,\ldots,N\!-\!1$:
\eqarr{
\label{s:hr}
s^{\mt{N\!-\!1}}\hs{-1ex}&=&\hs{-1ex}
\mn{\sqrt{\frac{2(N\!-1)}{N}}}\Big[
\ms{(N\!-\!1)}|\aver{u}|\re(u_{\mt{N\!-\!1}})
-|\aver{w}|\re(w_N)
\Big]\,,\quad\\
\label{s:hr-1}
s^{\mt{N\!-\!2}}\hs{-1ex}\hs{0ex}&=&\hs{-1ex} 
-\mn{\sqrt{\frac{2(N\!-2)}{N\!-1}}}|\aver{u}|\re(u_{\mt{N\!-\!1}})\,.
}

\item $T^i=\mathcal{S}_{ab}$, $a=1,\ldots,N\!-\!1$, $b=N$, $s^\mu(\Phi)\rightarrow
s_+^{ab}(\Phi)$:
\eqarr{
\label{s:aN:+}
s_+^{a,N}&=&|\aver{w}|\re(w_{a})+\delta_{a,{\mt{N\!-\!1}}}|\aver{u}|\re(u_{N})\,,\\
\label{s:ar:+}
s_+^{a,{\mt{N\!-\!1}}}&=&|\aver{u}|\re(u_{a})\,,~~a<N\!-\!1.
}

\item $T^i=\mathcal{A}_{ab}$, $a=1,\ldots,N\!-\!1$, $b=N$, $s^\mu(\Phi)\rightarrow
s_-^{ab}(\Phi)$:
\eqarr{
\label{s:aN:-}
s_-^{a,N}&=& -|\aver{w}|\im(w_{a})+\delta_{a,{\mt{N\!-\!1}}}|\aver{u}|\im(u_{N})\,,\\
\label{s:ar:-}
s_-^{a,{\mt{N\!-\!1}}}&=& -|\aver{u}|\im(u_{a})\,,~~a<N\!-\!1.
}
\end{itemize}
From Eqs.\,\eqref{MiN=0} and \eqref{Mir=0}, the fields $u_N,\,u_{\mt{N\!-\!1}},
\,w_N,\,w_{\mt{N\!-\!1}}$
are absent in the first term of the quadratic part of the potential in
Eq.\,\eqref{V:2}. The fields $\im w_N$ and $\im u_{\mt{N\!-\!1}}$ are also absent in
$s^\mu(\Phi)$ which make them massless. The real components $\re w_N$ and $\re
u_{\mt{N\!-\!1}}$ are present in Eqs.\,\eqref{s:0}--\eqref{s:hr-1}. The reminder
of the fields involving $w_{\mt{N\!-\!1}}$ and $u_N$ are only present in the
combinations of Eqs.\,\eqref{s:aN:+} and \eqref{s:aN:-} for $a=N\!-\!1$. The
orthogonal combinations shown in Eqs.\,\eqref{R} and \eqref{I} are then
massless.

%%%%%%%%%%%%%%%%%%%%%%%%%%%%%%%%%%%%%%%%%%%
\acknowledgments
This work was supported by {\em Fundação de Amparo à Pesquisa do Estado de São
Paulo} (Fapesp).
The author would like to thank Prof. Juan Carlos Montero and Prof. Vicente
Pleitez for critical discussions.
% Conselho Nacional de Desenvolvimento Cient\'\i fico e Tecnol\'ogico (CNPq).
%%%%%%%%%%%%%%%%%%%%%%%%%%%%%%%%%%%%%%%%%%%

%%%%%%%%%%%%%%%%%%%%%%%%%%%%%%%%%%%%%%%%%%%%%%%%%

\begin{thebibliography}{99}
\bibitem{endnote01}
For a CP invariant potential, in the real basis, we can count $N(N+1)/2$
(quadratic) $N^2(N^2+3)/4$ (quartic) parameters. If we take the $SO(N)_H$
reparametrization freedom into account, the potential can be parametrized by a
total of $N+N^2(N^2+3)/4$ real parameters.
\bibitem{endnote02}
Note that, for neutral solutions, $\langle \mathbb{M} \rangle=0$ means that all
charged scalars, i.e., one charged Goldstone and $N-1$ physical charged Higgs
bosons, are massless.
\bibitem{endnote1}
This name is justified because in this basis, when EM symmetry is preserved,
the components $u_a$, $a=1,\ldots,\ms{N\!-\!1}$ are already the physical charged
Higgs bosons.
\end{thebibliography}
\end{document}